\documentstyle[preprint,aps]{revtex}
\begin{document}
\baselineskip=16pt 

\begin{center}
\hfill    YITP-01-71

{\large {\bf  Effective vector-field theory and long-wavelength
 universality\\ of the fractional quantum Hall effect}}
\bigskip

{ K. Shizuya}
\bigskip

{\sl Yukawa Institute for Theoretical Physics\\
 Kyoto University,~Kyoto 606-8502,~Japan }
\end{center}

\noindent
{\bf Abstract} 
\par
\baselineskip=13pt 
{\small We report on an effective vector-field theory of the
fractional quantum Hall effect that takes into account
projection to Landau levels. 
The effective theory refers to neither the
composite-boson nor composite-fermion picture, but properly
reproduces the results consistent with them, thus revealing the
universality of the long-wavelength characteristics of the
quantum Hall states. In particular, the dual-field Lagrangian
of Lee and Zhang is obtained, and an argument is given
to verify the identification by Goldhaber and Jain
of a composite fermion as a dressed electron. The
generalization to double-layer systems is also 
remarked on.}\\

\smallskip


\baselineskip=14pt 
\noindent
{\bf 1. Introduction}
\smallskip

The early studies of the fractional quantum Hall effect 
(FQHE)  based on Laughlin's variational wave functions
[1-4] gave impetus to descriptions of the FQHE in
terms of electron--flux composites.
The Chern-Simons (CS) theories [5-7] realize the
composite-boson and composite-fermion pictures of the FQHE and
have been successful in describing the long-wavelength
characteristics of the fractional quantum Hall states. 
The CS approaches, however, make no explicit use of projection
to the lowest Landau level, a procedure which is crucial in the
wave-function approach.  
One might naturally wonder if and how the former are
compatible with the Landau-level structure and quenching of the
electronic kinetic energy.  The present paper is motivated to
answer this question affirmatively.

In this paper we wish to report on an effective vector-field
theory of the FQHE that takes into account projection to the
Landau levels [8]. The effective theory, constructed via
the electromagnetic responses and bosonization, does not refer
to either the composite-boson or composite-fermion picture, but
properly reproduces the results consistent with them, thus
revealing the universality of the long-wavelength
characteristics of the quantum Hall states. 
In particular, the dual-field Lagrangian of Lee and Zhang [6] 
is obtained. An argument is given to substantiate the
identification by Goldhaber and Jain [9] of the composite
fermion as a dressed electron. \\

%The generalization of our approach to double-layer quantum Hall
%systems is also discussed.\\

\noindent
{\bf 2. Electromagnetic response of Hall electrons}
\smallskip

Consider electrons in a plane with a perpendicular
magnetic field $B>0$, and study how they respond to 
weak external potentials $A_{\mu}(x)=(A_{0},{\bf A})$. 
The eigenstates of a freely orbiting electron are Landau levels
of energy $\omega_{\rm c} (n+{\scriptstyle {1\over2}})$ with 
$\omega_{c} \equiv eB/m^{*}$ and $n = 0,1,\cdots$.
This level structure is modified in the
presence of  $A_{\mu}(x)$, and the effect of level mixing it causes 
is calculated by diagonalizing the Hamiltonian $H$ with respect
to the true levels $\{n\}$.

Suppose now that an incompressible many-body state of uniform
density $\bar{\rho}$ is formed via the Coulomb
interaction (for $A_{\mu}=0$). The projected Hamiltonian then serves to
determine its response to weak electromagnetic potentials
$A_{\mu}(x)$, with the result 
\begin{eqnarray}
 L[A;\bar{\rho}]= \bar{\rho}&&\Big[-A_{0}
-\ell^{2}\,{1\over2}\, 
A_{\mu} \epsilon^{\mu \nu\lambda}\partial_{\nu}A_{\lambda} 
%\nonumber\\&&
+{\ell^{2}\over{2\omega_{\rm c}}} (A_{k0})^{2} -
{\sigma(\nu)\,\ell^{2}\over{2m^{*}}}  (A_{12})^{2}\Big]  +
O(\partial^{3}),
\end{eqnarray}
where  $\ell \equiv 1/\sqrt{eB},  A_{\mu\nu}\equiv
\partial_{\mu}A_{\nu}-\partial_{\nu}A_{\mu}$ and the electric
charge $e>0$ has been suppressed by rescaling
$eA_{\mu}\rightarrow A_{\mu}$; 
$\epsilon^{\mu\nu\rho}$ is a totally-antisymmetric tensor with
$\epsilon^{012} =1$ and $\sigma(\nu)$ is some factor of
$O(1)$, whose value is not needed in what follows. \\

\noindent
{\bf 3. Bosonization and effective theory}
\smallskip

The electromagnetic response we have calculated is essentially
the partition function 
$W[A] = \exp( i\int d^{3}x\, L[A;\bar{\rho}])$, 
which  is  written as a
functional integral over the electron fields.
Once such $W[A]$ is known it is possible to reconstruct it
through the quantum fluctuations of a boson field. 
The rules of functional bosonization [10] tell us that 
the response $W[A]$ is reconstructed from a bosonic
theory of a 3-vector field 
$b_{\mu}=(b_{0},b_{1}, b_{2})$, with the action $Z[b]$
given by Fourier transforming (given) $W[v]$,
\begin{equation}
e^{iZ[b]} \equiv \int [dv_{\lambda}]\,e^{ i\int d^{3}x\, 
b_{\mu}\epsilon^{\mu \nu\rho} \partial_{\nu}v_{\rho} }\, W[v].
\label{Seffb}
\end{equation}
The  $b_{\mu}$ field is coupled to $A_{\mu}$ through the coupling 
$-A_{\mu} \epsilon^{\mu \nu\lambda} \partial_{\nu}b_{\lambda}$.

Let us now substitute $L[A;\bar{\rho}]$ into Eq.~(\ref{Seffb})
and construct an equivalent bosonic theory that recovers it.
The relevant functional integration is best carried out in the
following way:  First make a shift 
$v_{\mu}=v'_{\mu} +f_{\mu}[b]$ and choose
$f_{\mu}[b]$ so as to decouple $v'_{\mu}$ and $b_{\mu}$ in the
exponent. All the $b_{\mu}$ dependence is then isolated
in the background piece $f_{\mu }= (1/\bar{\rho}\ell^{2})\,
b_{\mu} + O(\partial\,b)$, leading to the action 
$Z[b]=\int d^{3}x\,L^{(0)}[b]$ with 
\begin{eqnarray}
L^{(0)}[b]= &&-{1\over{\ell^{2}}}\,b_{0}+
{\pi\over{\nu}}\,b_{\mu} \epsilon^{\mu\nu\lambda}
\partial_{\nu}  b_{\lambda} 
%\nonumber\\&&
+{\pi\over{\nu\,\omega}}\,(b_{k0})^{2} - 
{\sigma(\nu)\,\pi\over{\nu\,m^{*}}}\,(b_{12})^{2}  
+ O(\partial^{3}) ,
\label{Lzerob}
\end{eqnarray}
where the filling factor $\nu= 2\pi \ell^{2}\bar{\rho}$ and   
$b_{\mu \nu}=\partial_{\mu}b_{\nu} -\partial_{\nu}b_{\mu}$.

An advantage of the bosonic theory is that the Coulomb interaction is
handled exactly. It further admits inclusion of new degrees of
freedom, vortices, which describe Laughlin's quasiparticle excitations
over the FQH states. 
Eventually the bosonic theory that takes into account
both the Coulomb interaction and vortices is described by the
Lagrangian
\begin{equation}
L_{\rm eff}[b]= -A_{\mu} \epsilon^{\mu\nu\rho}
\partial_{\nu}b_{\rho} 
-2\pi \tilde{j}_{\mu}b_{\mu}  + L^{(0)}[b] 
-{1\over{2}}\, \delta b_{12}\,V\, \delta b_{12} ,
\label{Leffb}
\end{equation}
where $\tilde{j}_{\mu}$ stands for the vortex 3-current;
$\delta b_{12}\,V\, \delta b_{12} \equiv 
\int d^{2}{\bf y}\, \delta b_{12}(x)\,
V({\bf x -y})\, \delta b_{12}(y)$ denotes the (bosonized)
Coulomb interaction and $\delta b_{12}= b_{12} -\bar{\rho}$. 
Upon quantization the CS term $b\epsilon\partial b$ in
$L^{(0)}[b]$ combines with the kinetic term $(b_{k0})^{2}$ to
yield a ``mass'' gap $\omega$ while the $(b_{12})^{2}$ term
yields only a tiny $O(\nabla^{2})$ correction to it.

This $L_{\rm eff}[b]$ almost precisely agrees with the dual-field
Lagrangian of Lee and Zhang  (LZ) [6], derived within the
Chern-Simons-Landau-Ginzburg (CSLG) theory.
The only difference to $O(\partial^{2})$ lies in the $(b_{12})^{2}$
term, which, however, is unimportant at long wavelengths.

The fact that we have practically arrived at the LZ dual
Lagrangian  without invoking the composite-boson picture would
thus reveal the following: 
(1) The long-wavelength characteristics of the FQH states, as
governed by the LZ dual Lagrangian, are determined universally
by the filling factor and some single-electron characteristics,
independent of the details of the FQH states.  
(2) In the CSLG theory the random-phase approximation
properly recovers the effect of the Landau-level structure,
crucial for determining the electromagnetic response.

It is instructive to read off the vortex charge from
the $-2\pi \tilde{j}_{\mu}\, b_{\mu}$ coupling in $L_{\rm eff}[b]$.  
The $b_{\mu}$ field acquires, in the presence of the electromagnetic
coupling 
$-A_{\mu} \epsilon^{\mu \nu\rho} \partial_{\nu}b_{\rho}$, 
a background component, i.e.,
$b_{\mu}=b'_{\mu} +f_{\mu}$ with 
$f_{\mu}= (\nu/2\pi)\, A_{\mu} + O(\partial A)$, so that 
\begin{equation}
-2\pi \tilde{j}_{\mu}\,b_{\mu}= - \nu\, 
\tilde{j}_{\mu}A_{\mu} + \cdots,
\end{equation}
which reveals the vortex charge of $-\nu e$. \\

\noindent
{\bf 4. Composite fermions}
\smallskip

The fermionic CS theory [7] realizes the composite-fermion
description of the FQHE. There the composite fermions are
visualized as electrons carrying an even number $2p$ of flux
quanta of the CS field $c_{\mu}(x)$.   
The fractional quantum Hall states at the principal filling fractions 
$\nu \equiv 2\pi\, \ell^{2}\,\bar{\rho}
=\nu_{\rm eff}/(2p \nu_{\rm eff} \pm1)$
are thereby mapped into an integer quantum Hall state of composite
fermions in the reduced field
$B_{\rm eff} = (1- 2p\,\nu)B$ at $\nu_{\rm eff}=1,2,\cdots$.

It is possible to cast, via the electromagnetic response of the
composite fermion, this fermionic CS theory into an equivalent
bosonic version, with the Lagrangian
\begin{eqnarray}
L^{\rm CF}_{\rm eff}[b,c]= &&-e(A_{\mu} + c_{\mu}) 
\epsilon^{\mu\nu\rho} \partial_{\nu}b_{\rho} 
-2\pi \tilde{j}_{\mu}b_{\mu} 
%\nonumber\\&& 
+ L^{(0)}[b\,; B_{\rm eff}] + L_{\rm CS}[c],
\label{LCFb}
\end{eqnarray}
where $L^{(0)}[b; B_{\rm eff}]$ stands for $L^{(0)}[b]$ of
Eq.~(\ref{Lzerob}) with
$B \rightarrow B_{\rm eff}$ and  
$\nu \rightarrow \nu_{\rm eff}$.

Let us here try to eliminate $c_{\mu}$ 
from $L^{\rm CF}_{\rm eff}[b,c]$ and derive an equivalent theory 
of the $b_{\mu}$ field alone.  
We omit the detail here. The result precisely coincides with
$L_{\rm eff}[b]$ in Eq.~(\ref{Leffb}), apart from (unimportant)
$(b_{12})^{2}$ terms. 
This verifies, in particular, that the composite-boson and
composite-fermion descriptions of the FQHE as well as our
(projection+bosonization) approach are perfectly consistent 
in the long-wavelength regime.

As for the nature of the composite fermion Goldhaber and
Jain~[9] identified it as a dressed electron with
bare charge $-e$, and argued on the basis of 
Laughlin's wave functions that the bare charge is screened in
the composite-fermion medium to yield local charge $-\nu e$.

It is possible to substantiate such characteristics of
composite fermions within the CS approach.
First, applying Laughlin's reasoning reveals that the vortex
excitations over the composite-fermion ground state are nothing
but the composite fermions (or holes) themselves. 
Counting of the vortex charges [or simply Eq.~(6)] shows that each
composite fermion has ``bare'' charge $-e$ in response to the "bare"
field $A_{\mu} +c_{\mu}$.  
Note next that, upon integration over $b_{\mu}$ with 
$L^{\rm CF}[b,c]$, the CS field $c_{\mu}$ acquires a background piece
proportional to $A_{\mu}$, leading to substantial dressing (or
renormalization) of the bare field
\begin{equation}
A_{\mu}+c_{\mu} = \pm\,
(\nu/\nu_{\rm eff})\,A_{\mu} +\cdots,
\end{equation}
which shows that a composite fermion has fractional charge 
\begin{equation}
Q_{\rm CF}=-(\nu/\nu_{\rm eff})\, e 
= - e/(2p\nu_{\rm eff} \pm1) 
\end{equation}
 in response to $A_{\mu}$. (In the above the $\pm$ sign refers
to the sign of $B_{\rm eff}/B$.)  
Hence the bare charge $-e$ coupled to $A_{\mu}+c_{\mu}$ is the same as
the renormalized charge $Q_{\rm CF}$ probed with $A_{\mu}$. This
special dressing of the bare field has a natural consequence: 
The composite fermion with bare charge $-e$ feels an effective
electric field $E^{A+c}_{y}\equiv E_{y}+c_{20}=  
\pm (\nu/\nu_{\rm eff}) E_{y} +\cdots$ in the effective
magnetic field
$B_{\rm eff}= \pm (\nu /\nu_{\rm eff})B$ so that 
it drifts with the same velocity as the electron, as it should,
\begin{equation} 
v_{x}^{\rm drift}=E^{A+c}_{y}/B_{\rm eff}=E_{y}/B.
\end{equation}

%\newpage

\noindent
{\bf 5. Concluding remarks}
\medskip

In this paper we have presented an effective vector-field theory of
the FQHE. It does not refer to either the composite-boson or
composite-fermion picture, and simply supposes an incompressible FQH
ground state of uniform density. 
Our approach by itself does not tell at which filling fraction
$\nu$ such a  state is realized. Instead, it tells us that once
such an incompressible state is formed its long-wavelength
characteristics are fixed universally, independent of the
composite-boson and composite-fermion pictures. In this sense,
our approach is complementary to the picture-specific CS approaches.

The CS approaches without Landau-level projection encounter some
subtle difficulties [11], when generalized to double-layer
systems where intra-Landau-level collective excitations play an
important role in the low-energy regime. In our approach it
turns out possible to construct an effective theory that
properly incorporates the effect of the dipole-active
inter-layer out-of-phase collective mode (via its
electromagnetic response in the single-mode approximation)
[12]. Details will be reported elsewhere. \\

\noindent
{\bf Acknowledgments}
\medskip

This work is supported in part by a Grant-in-Aid for 
Scientific Research from the Ministry of Education of Japan, 
Science and Culture (No. 10640265).  \\

%\newpage
\noindent
{\bf References}\\

\noindent
[1] R. B. Laughlin, Phys. Rev. Lett. 50 (1983) 1395.

\noindent
[2] S. M. Girvin and A. H. MacDonald,  Phys. Rev. Lett. 58 
(1987) 1252 .  

\noindent
[3] N. Read, Phys. Rev. Lett. 62 (1989) 86. 

\noindent
[4] J.~K.~Jain, Phys. Rev. Lett. 63 (1989) 199.

\noindent
[5] S. C. Zhang, T.H. Hansson, and S. Kivelson, Phys. Rev. Lett. 
62 (1989) 82.   

\noindent
[6] D.-H. Lee and S.-C. Zhang, Phys. Rev. Lett. 66 
(1991) 1220. 

\noindent
[7] A. Lopez and E. Fradkin, Phys. Rev. B 44 (1991) 5246.

\noindent
[8] K. Shizuya, Phys. Rev. B 63 (2001) 245301 1-8.

\noindent
[9] A.~S.~Goldhaber and J.~K.~Jain, Phys. Lett. A 199 
(1995) 267.
 
\noindent
[10] F. A. Schaposnik, Phys. Lett. B 356 (1995) 39, and
earlier references cited therein. 

\noindent
[11] A. H. MacDonald and S.-C. Zhang, Phys. Rev. B 49
(1994) 17 208 .  

\noindent
[12] K. Shizuya, in preparation.

\end{document}